\documentclass[twocolumn,showpacs,preprintnumbers,amsmath,amssymb]{revtex4-1}
\usepackage{graphicx}
\usepackage{amsmath}
 \usepackage{epstopdf}
\usepackage{color}

\usepackage{color}

\begin{document}
\title{Wind generated rogue waves in an annular wave flume}

\author{A. Toffoli$^{1}$}
\author{D. Proment$^2$}
\author{H. Salman$^{2}$}
\author{J. Monbaliu$^{3}$}
\author{F. Frascoli$^4$}
\author{M. Dafilis$^5$}
\author{E. Stramignoni$^6$}
\author{R. Forza$^6$}
\author{M. Manfrin$^6$}
\author{M. Onorato$^{6,7}$}

\affiliation{
$^1$Department of Infrastructure Engineering, The University of Melbourne, Parkville, VIC 3010, Australia;\\
$^2$School of Mathematics, University of East Anglia, Norwich Research Park, Norwich, NR4 7TJ, UK;\\
$^3$K.U. Leuven, Kasteelpark Arenberg 40, 3001 Heverlee, Belgium;\\
$^4$Department of Mathematics, Faculty of Science, Engineering and Technology, Swinburne University of Technology, Hawthorn, VIC 3122, Australia;\\
$^5$Department of Health and Medical Sciences, Faculty of Health, Art and Design, Swinburne University of Technology, Hawthorn, VIC 3122, Australia;\\
$^6$Dipartimento di Fisica, Universit{\`a} degli Studi di Torino, Via Pietro Giuria 1, 10125 Torino, Italy;\\
$^7$INFN, Sezione di Torino, Via Pietro Giuria 1, 10125 Torino, Italy.
}

\date{\today}

\begin{abstract}
We investigate experimentally the statistical properties of a wind-generated wave field and the spontaneous formation of rogue waves in an annular flume. Unlike many experiments on rogue waves where waves are mechanically generated, here the wave field is forced naturally by wind as it is in the ocean. What is unique about the present experiment is that 
the annular geometry of the tank makes waves propagating circularly in an {\it unlimited-fetch} condition. Within this peculiar framework, we discuss the temporal evolution of the statistical properties of the surface elevation. We show that rogue waves and heavy-tail statistics may develop naturally during the growth of the waves just before the wave height reaches a stationary condition. 
Our results shed new light on the formation of rogue waves in a natural environment.   
\end{abstract}
\maketitle

Rogue waves are rare events of exceptional height  that may surge without warnings \cite{akhmediev09b,dysthe08,onorato2013review,erkintalo2015rogue}. 
This peculiar phenomenon is ubiquitous. It has been observed in different contexts such as gravity and capillary waves \cite{ONO05,ONO09,shats2010capillary,chabchoub2011rogue,chabchoub2012super,toffoli2013pre}, optical fibres \cite{solli07,montina2009non,kibler2010peregrine,kibler2015superregular,walczak2015optical,akhmediev2016roadmap,dudley2014instabilities,Suret:16,narhi2016real}, superfluid helium \cite{ganshin2008observation} and plasmas \cite{bailung2011observation,tsai2016generation}.  Because of their universal  and potentially detrimental nature, there is a pressing need to understand their physics in order to predict and control them.

The generating mechanisms can be disparate \cite{kharif09}. These include the spatio-temporal linear focussing of wave energy \cite{PEL,pisarchik2011rogue}, the focussing due to bathymetry and currents (see e.g. \cite{white1998chance,brown2001space,heller2008refraction}) and the self-focussing that results from the Benjamin-Feir instability \cite{zakharov2009}.
The latter is described by exact breather solutions of the nonlinear Schr\"odinger (NLS) equation \cite{akhmediev87}, which are coherent structures that oscillate in space and/or time. Interestingly enough, breathers can also exist embedded in random waves \citep{onorato01}. Provided that the ratio of the dominant wave steepness to the spectral bandwidth is $O(1)$ and propagation is unidirectional, large amplitude structures can occur often enough to originate strong deviations from Gaussian statistics \cite{onorato01,janssen03,ONO06,ONO09,walczak2015optical}. Therefore, breathers have been considered in various fields of physics as a plausible prototype of rogue waves.

Such solutions have been reproduced experimentally in wave tanks using prescribed boundary conditions at the wave maker \cite{chabchoub2011rogue}.  Indeed, the standard form of NLS equation describes the nonlinear dynamics of a pre-existing (initial) wave field, which propagates without gaining or losing energy. This framework, however, is not transferable in a straightforward manner to systems driven by external forcing. The most obvious example of such a context is the ocean, where the oscillatory motion of the water surface is generated by the forcing of local wind (the resulting wave field is generally known as wind sea). Waves then grow with fetch and/or time until a quasi-stationary condition is reached, i.e. a fully developed sea \cite{komen84}.
Experimental work in wave tanks where waves are generated only by winds have been reported in the past, see for example Ref.~\cite{huang1980experimental,zavadsky2013statistical,caulliez2012higher}. 
Due to finite-length constraints of wind-wave flumes, experiments are performed in 
fetch-limited and statistically stationary conditions, with moderately small fetches. Under these circumstances, it has been observed that statistical properties of the surface elevation only weakly deviates from Gaussian statistics. 

In the present Letter, we discuss a laboratory wind-sea experiment in an annular  flume, over which a constant and quasi-homogeneous wind blows. Instead of the {\it fetch-limited} and time-independent settings that have characterised previous experiments in rectilinear flumes, the annular geometry impose a so-called 
{\it duration-limited} condition \cite{young1999wind}.  
Note that in the case of an unforced and undamped wave system, the dynamics in space and time are related to the leading order in nonlinearity and dispersion by the group velocity.
In the presence of wind forcing,  the relation between the temporal and spatial dynamics is not trivial, see \cite{stiassnie2007temporal}.

 Our peculiar facility allows the observation of the continuous growth in time, from the initial still water surface, to the fully developed condition. 
We show that during the very early stages of the generation, characterised by a growth of the wave height and a downshift of the spectral peak, the statistics is close to Gaussian. Just before the wave spectrum reaches its stationary state, the maximum deviations from Gaussian statistics and formation of rogue waves are observed. Once stationarity is reached, the statistics falls back to a Gaussian regime. 

\begin{figure}[]
\includegraphics[width=8.7cm]{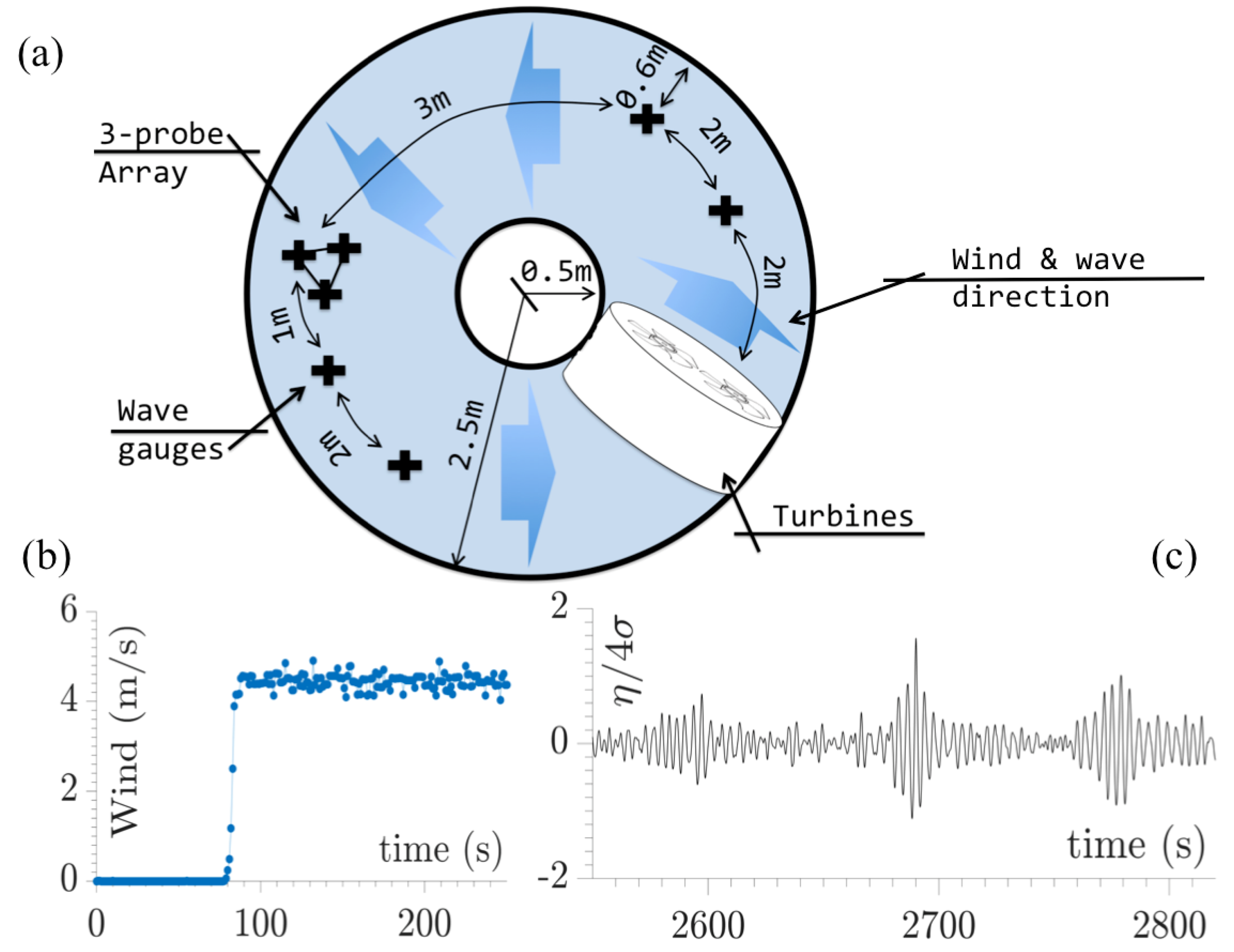}
\caption{Experimental set up (not in scale, panel a); example of wind speed (panel b); and example of water surface elevation (normalised by four times the standard deviation of the 10-minute record), including a rogue wave with wave height 2.7 times higher than the significant wave height (panel c).}\label{tank}
\end{figure}

The experiment was conducted in the geophysical circular wave flume at the University of Turin. The flume has an outside diameter of 5\,$m$ and an inside diameter of 1\,$m$ (Fig.~\ref{tank}a). The annular region of 2\,$m$  width was filled with 0.46 $m$ of water,  leaving a closed air chamber above the water surface  of approximately 0.5\,$m$. Two 2.2\,$KW$ industrial fans (flow rate of 9600\,$m^3/h$) were then mounted in the circuit for the generation of the wind. The air flow was measured by a three-dimensional ultrasonic anemometer, which operated at a sampling rate of 20.8\,$Hz$, and a hot wire, which recorded the air flow at sampling frequency of 1\,$Hz$. Both instruments were deployed at about 0.3\,$m$ above the still water level. The water surface was traced by a total of seven capacitance wave gauges, operating at a sampling frequency of 50\,$Hz$. Four wave gauges were deployed at  a distance of 2\,$m$, 4\,$m$, 8\,$m$ and 10\,$m$ from the turbines (distances are taken counter-clockwise along the arc-length). Without loss of generality, only wave data from the farthest probes (at 8\,$m$ and 10\,$m$ from the turbines) are discussed herein. An additional three-gauge array was installed at about 7\,$m$ from the fans. The array had a shape of an equilateral triangle circumscribed in a circle of diameter of 0.2\,$m$. The configuration of this array was specifically designed to measure the full directional spectrum. A high-resolution acoustic velocimeter was also placed at about 0.2 $m$ below the surface at rest  to measure the water velocity.
Note that the velocity field in the water comprises a wave-induced oscillatory motion and a wind-induced current.
%
%

The still water surface was the initial condition for the experiment. Fans were then turned on to produce a steady wind that reached rapidly a target speed of 4\,$m/s$ (see an example of wind time series as recorded by the hot wire in Fig.~\ref{tank}b). Wind was kept blowing without interruptions for two hours. After such time, the air friction velocity was calculated to be $u_* = 0.21$\,$m/s$ and the wind-induced water velocity was measured to be approximately $U=0.07\:m/s$. The water surface elevation was monitored continuously during the entire test.  The same experimental test was repeated four times to increase the statistical robustness of the results. For each experiment, the time series of the surface were subdivided into 10 minute records, and post processing was then carried out.

\begin{figure}[]
\includegraphics[width=8.5cm]{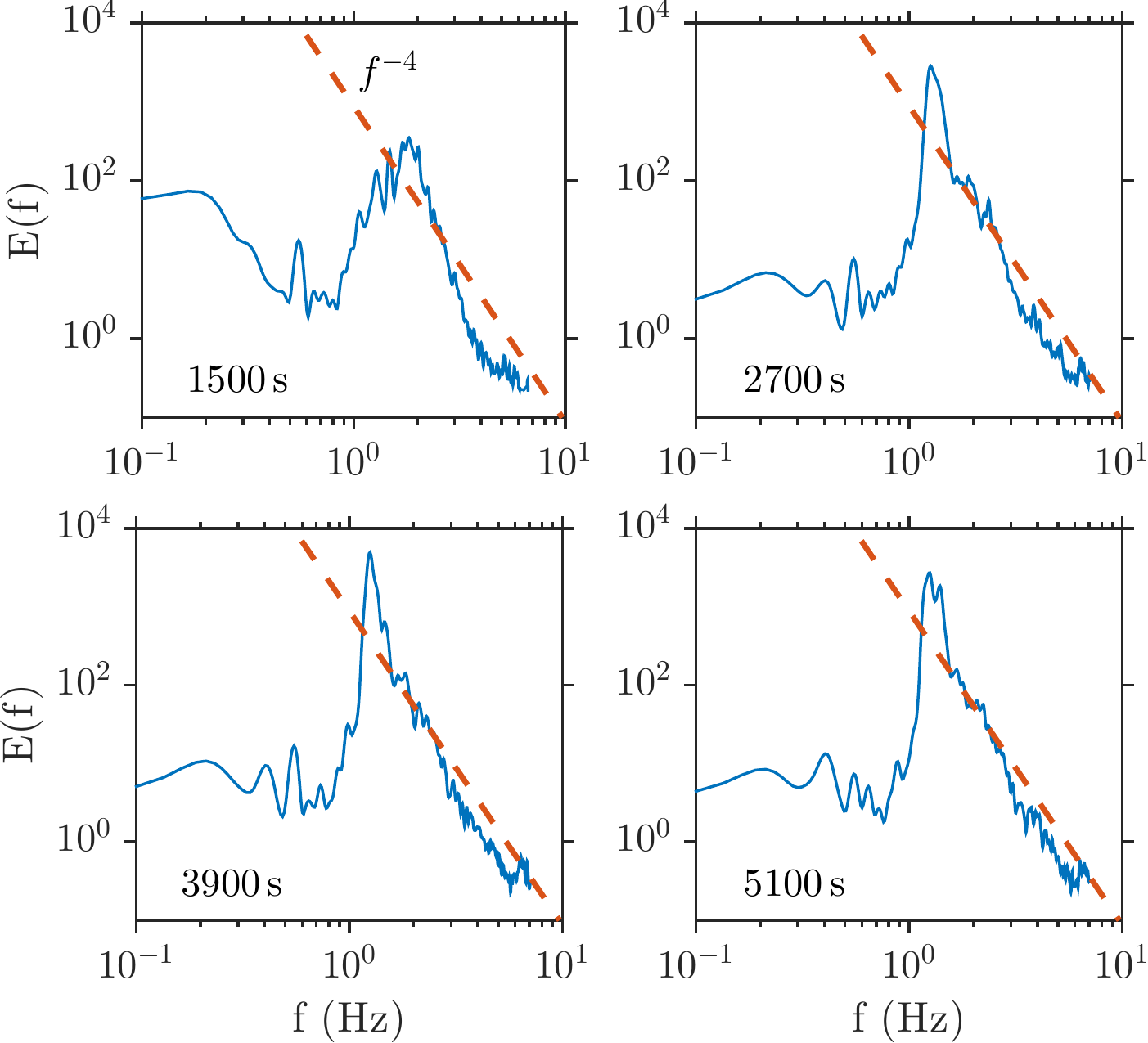}
\caption{Temporal evolution of the wave spectrum as a function of the intrinsic frequency. 
To  guide the eye, a power law $f^{-4}$, corresponding to the prediction of the wave turbulence theory is also shown as a dashed line. } \label{spec1d}
\end{figure}

A Fourier Transform algorithm was applied to reconstruct the distribution of the wave energy in frequency domain. For each 10-minute block, the spectrum was calculated from non-overlapping windows of about 41\,$s$ (i.e. 2048 data points) and then averaged (averaging over the different realizations was also performed).
The spectra at different time intervals are
shown in Fig.~\ref{spec1d}. It is interesting to note the development of a power law spectrum and the shift of the peak of the spectrum towards lower frequencies in time. Both these effects are an evidence of a nonlinear transfer of energy during the wind sea evolution. A power law $f^{-4}$ predicted by the Weak Wave Turbulence theory \cite{zakharov1967energy} is also plotted in the figure. As also observed in many wave tank experiments and in the ocean, the spectra in the stationary regime appear to be somehow steeper than the theoretical predictions \cite{deike2015role,denissenko2007gravity}.

\begin{figure}[]
\includegraphics[width=8.5cm]{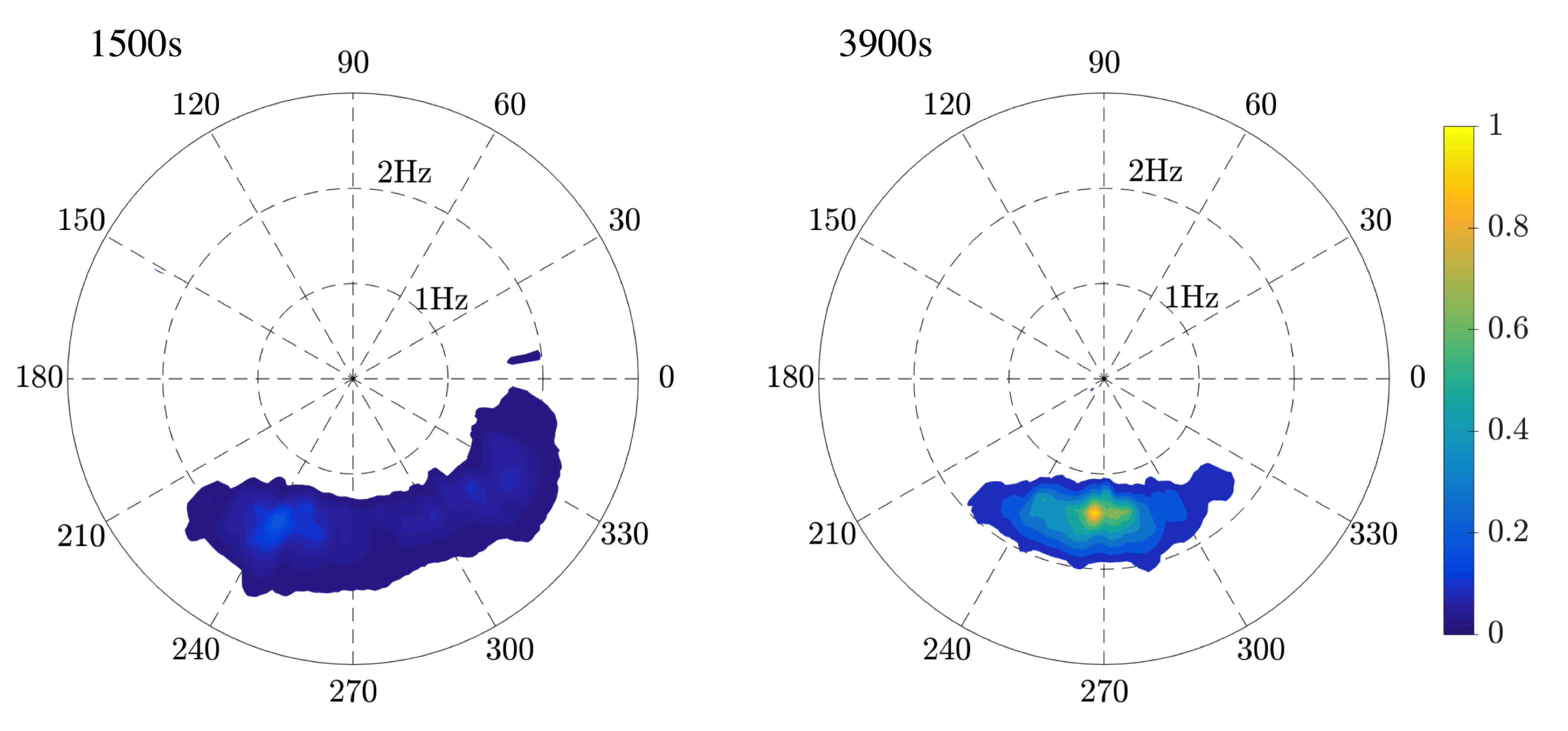}
\caption{Directional wave spectra in polar coordinates $(f,\theta)$ at 1500 and 3900 $s$.  Concentric lines indicate frequency of 1 and 2 $Hz$  from inside to outside. The colour scale indicates the energy density normalized with the maximum of the energy density at time 3900 $s$. The angles  $\theta=0^\circ$ and $180^\circ$ correspond to waves travelling radially, inward and outwards relative to the center of  the annular wave flume, respectively; $\theta=270^{\circ}$ and $\theta=90^{\circ}$ corresponds to waves travelling tangentially, clockwise and anti-clockwise, respectively.}\label{spec2d}
\end{figure}

The directional spectrum was computed with a wavelet directional method \cite{donelan96} from data recorded by the three-gauge array. In Fig.~\ref{spec2d} we show the  spectral energy density  as a function of the frequency and angle in polar coordinates;  the tangential direction corresponds to an angle of $270^{\circ}$.
%
%
%
%
The spectra highlight the fact that energy also spreads over angles, analogously to real ocean waves forced by the wind.   Note that the directional distribution is asymmetric in the directional domain due to a non-uniform cross-tank distribution of the wind speed. During the growth phase, energy moves toward lower frequencies, developing a rather narrow banded peak. At the same time, the energy concentrates over a narrower directional band (see righthand panel in Fig.~\ref{spec2d}).  

%
%
\begin{figure}[h]
\includegraphics[width=7cm]{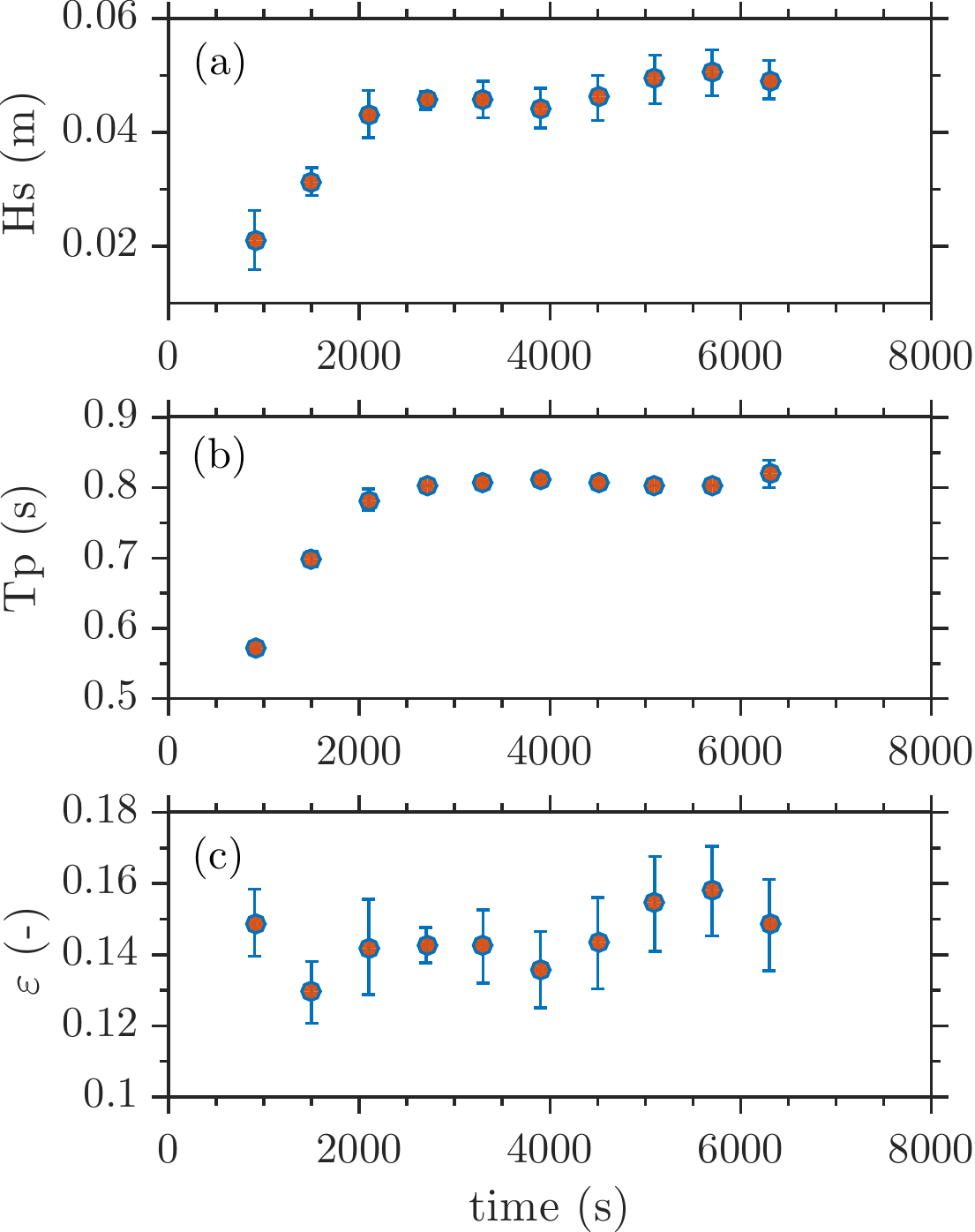}
\caption{Temporal variation of the significant wave height $H_s$ (a), peak period $T_p$ (b) and wave steepness $\varepsilon$ (c).}\label{fig:hs}
\end{figure} 

Using the wave spectra, it is possible to calculate the evolution in time of the significant wave height $H_s$ (i.e. four times the standard deviation of the surface elevation), the peak period and the steepness. The latter is a measure of the degree of nonlinearity of the system and is defined as $\varepsilon = k_p H_s / 2 $ with $k_p$ being the wavenumber at the spectral peak. Such quantities are displayed in  Fig.~\ref{fig:hs}. We recall that at time $t=0$ the surface is flat. As the wind starts, waves grow until they reach a quasi-stationary state characterised by a constant $H_s$ of about 0.048\,$m$ (after more or less half an hour). In oceanography such condition is usually referred to as ``\emph{fully developed condition}''.  Wave breaking was observed during the evolution. As observed directly from the spectra (Fig.~\ref{spec1d}), the peak period, $T_p$, also grows monotonically until a stationary state is reached (Fig.~\ref{fig:hs}b). The wave steepness remains steady and normally rather high ($\varepsilon = 0.145$, on average) throughout the experiments (i.e. during both the growing and fully developed stage).  
A number of causes may contribute to the formation of such stationary states.
The  phase velocity ($c_p$ = 1.25 $m/s$) becomes almost an order of magnitude larger than the friction velocity ($u^*$ = 0.21 m/s), meaning that energy transfer from the wind to the waves becomes smaller and smaller.
Moreover, in Ref. \cite{janssen07} an interesting analysis on the coupling coefficient of the Wave Kinetic Equation in arbitrary depth has been performed. The results show that in the deep water regime, $k_ph\rightarrow\infty$, the spectral peak is subjected to a downshifting phenomenon. However, when the peak wavenumber of the spectrum approaches the threshold value $k_ph$ = 1.36, the downshift vanishes. This is due to the fact that the coupling coefficient approaches zero as $k_ph \rightarrow$ 1.36. Such theoretical analysis was further confirmed by numerical simulations  (see  Fig. 4 in  \cite{janssen07}).  In our experiments waves start with deep water conditions, but the stationary state is reached at  an intermediate water depth with  $k_ph\sim$ 2 (modulational instability is still possible for such a value of $k_ph$).
 Under these circumstances,  the energy accumulates at the peak of the spectrum (where the forcing is located) until wave breaking takes place, i.e. a fast transfer of energy from low to high wave numbers, not related to four-wave resonant interactions. This mechanism allows for the formation of a stationary state characterized by an intermittent direct cascade due to wave breaking. Indeed, a significant amount of breaking was observed during the experiment in the stationary regime.
A concurrent cause of the formation of a stationary state can be related to the fact that  waves travel circularly around the centre of the annular region; once the stationary regime is reached ($f_p$ = 1.25 $Hz$), there are less than 10 waves in the  tank and  finite size effects may, therefore, influence the nonlinear transfer.

\begin{figure}[]
\includegraphics[width=8.5cm]{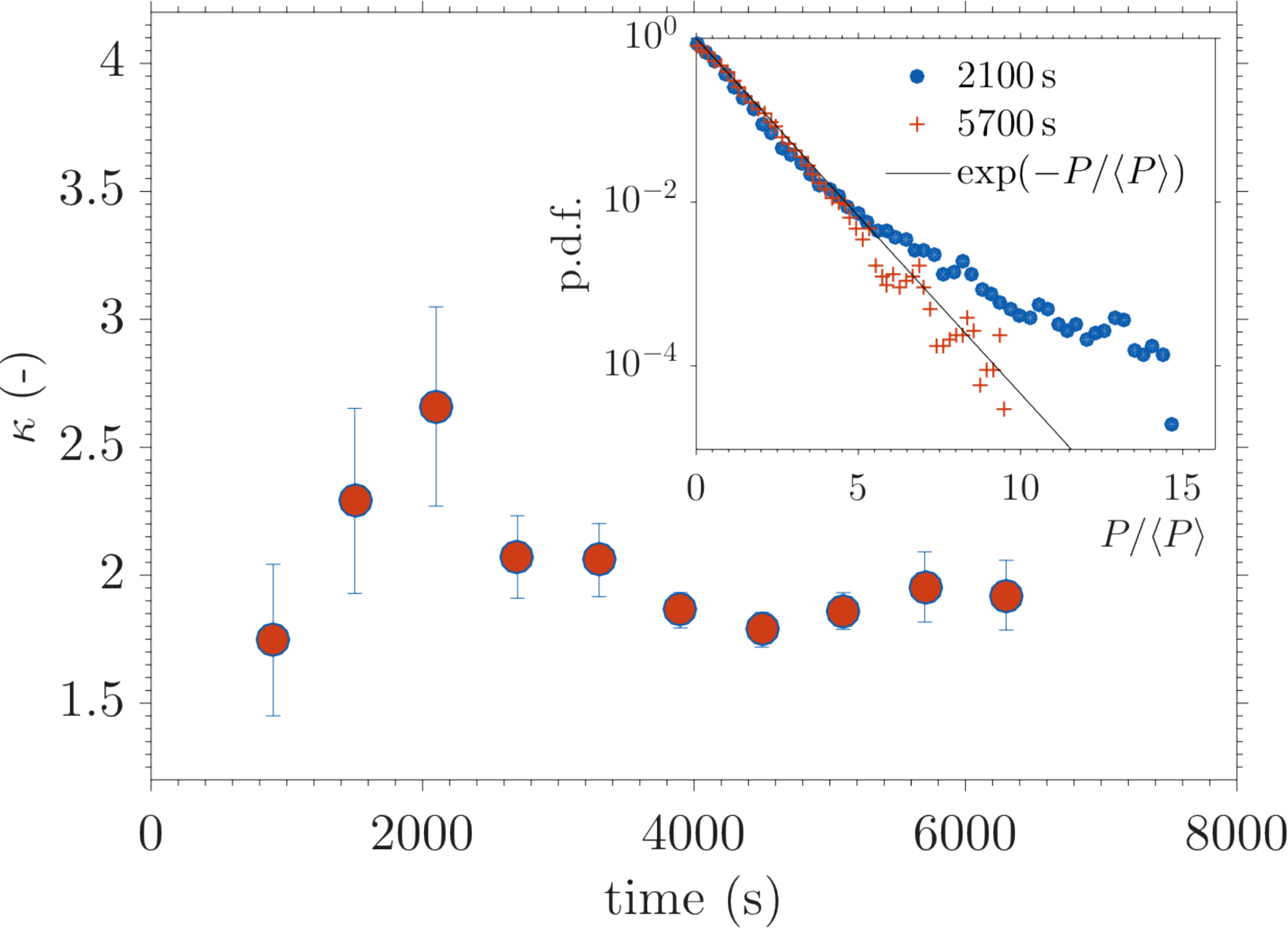}
\caption{Temporal evolution of $\kappa$ of the wave envelope (main panel) and probability density functions ($p.d.f.$) of the normalised wave intensity $P / \langle P \rangle$ (inset) at the time of maximum kurtosis (2100\,$s$) and at full development (5700\,$s$). The wave intensity is defined as the square modulus of the wave enveloped divided by its mean,  The $p.d.f.$ for a Gaussian random process, i.e. $\exp(-P / \langle P \rangle)$, is shown as reference.}\label{fig:kurt}
\end{figure}

In the physical space, the wave field is characterised by well defined packets, which are consistent with the narrow banded spectral peak; see an example of time series in Fig.~\ref{tank}c.  As can be seen from the figure, a rogue wave with height larger than 2.7 times $H_s$ is present in the time series. In order to investigate its origin and the statistical relevance of such waves we consider the normalized fourth-order moment, $\kappa$, of the probability density function ($p.d.f.$) of the wave envelope computed as
\begin{equation}
\kappa=\frac{N\sum_{i=1}^N |A_i|^4}{(\sum_{i=1}^N |A_i|^2)^2},
\end{equation}
where $|A_i|$ is the envelope of the time series of the surface elevation computed using the Hilbert transform over $N$ samples.
This quantity  allows us
 to verify whether such a rogue wave is a rare event of a Gaussian distribution or whether it belongs to a non-Gaussian distribution (see Fig.~\ref{fig:kurt}). Note that $\kappa$ is calculated after removing the bound modes, namely the components at frequencies greater than 1.5 and lower than 0.5 times the dominant frequency, which are primarily generated by second-order effects \cite{LH63}. In doing so, the nonlinear dynamics of free waves remains the only nonlinear mechanisms responsible for the formation of extreme events. 
%
%
%
If the sea state is a Gaussian random process, the kurtosis of the surface elevation is equal to 3 and the corresponding value of $\kappa$ (calculated on the envelope of the surface elevation) is equal to 2. 
During the wave growth, however, $\kappa$ clearly exhibits a monotonic increase until a maximum is reached after 2100\,s (main panel in Fig.~\ref{fig:kurt}). It is interesting to note that $\kappa$ reaches remarkably high values. This strongly non-Gaussian conditions are attained when wave energy focuses both in the frequency and directional domain. The deviation from Gaussianity is substantiated robustly by the heavy tail of the $p.d.f.$ of the wave intensity $P$, i.e. the square modulus of the wave envelope (see the inset in Fig.~\ref{fig:kurt}). For longer duration, $\kappa$ drops to the value of 2, at which it remains throughout the fully developed stage. Under these circumstances, the tail of the $p.d.f.$ of $I$ fits the one expected for a Gaussian random process, i.e. $\exp({-P})$, see inset in Fig.~\ref{fig:kurt}. This result is consistent with numerical simulations of the long-time evolution of the statistical moments of wind seas in \cite{annenkov2009evolution}, where the contribution of free wave nonlinear dynamics to wave statistics is shown to be negligible. 
Concerning the possible causes of the formation of the rogue waves, we can exclude with some confidence  that their formation
is the result of a simple superposition of linear waves; indeed, in this latter case, the probability density function of the intensity would have followed an exponential distribution. Moreover, in order to compute the envelope we have removed bound modes; this implies that the deviation of $\kappa$ from the linear prediction cannot be due to Stokes-like corrections. As mentioned, in our experiment  $k_p h$   is always larger than 1.36. This implies that modulational instability for incoherent wave systems is still a possible candidate for explaining the observed rogue waves. It is interesting to note that at the time of the formation of rogue waves, the spectrum experiences a fast change. This seems to be in accordance with results described in \cite{onorato2016origin,annenkov2016modelling} where changes of the kurtosis are associated with rapid changes of the spectrum.
It is also interesting to note that in our experiment we observe the formation of rogue waves in a nondimensional water depth of $k_ph\simeq 2$; this is exactly the range of nondimensional water depth at which both the celebrated Draupner and  Andrea waves were recorded in the North Sea \cite{bitner2014north}. Such depth may hide new physics that definitely needs more investigation.



In conclusion, we have presented a laboratory experiment in an annular wind-wave flume to study the statistical properties of wind-generated waves and rogue wave probability. The facility allows  the full evolution of the wave field, from its generation to the fully developed stage. As wind starts blowing, an erratic wave field is generated. Rogue waves are detected just before reaching a stationary state. Consequently, strong deviations from Gaussian statistics are observed. 
We are fully aware that the experimental model is not the ocean. Nonetheless, for the first time, large deviation from Gaussianity have been observed during the development of a wind-forced wave field.  To some extent, the condition of infinite fetch modelled in the present experiment exists in the Southern Ocean, where strong winds (the Roaring Forties, Furious Fifties and Screaming Sixties \cite{lundy2010godforsaken}) blow around the Antarctic continent. Waves in the Southern Ocean are indeed regarded to be the fiercest on the planet.      

{\bf Acknowledgments} Experiments were supported by the European Community Framework Programme 7, European High Performance Infrastructures in Turbulence (EuHIT), Contract No. 312778. M.O. acknowledges Dr B. Giulinico and Dr G. Di Cicca for interesting discussions. A. Iafrati is acknowledged for discussions and for providing the wave gauges from INSEAN.
\bibliography{references}
\end{document}